\documentclass[onecolumn,showpacs,preprintnumbers,amsmath,amssymb,floatfix]{revtex4}
\usepackage{graphicx}
\usepackage{dcolumn}
\usepackage{bm}
\begin{document}

\preprint{PRD}

\title[]{Stochastic background from inspiralling double neutron stars}

\author{Tania Regimbau}
\affiliation{Dpt. ARTEMIS \\Observatoire de la C\^ote d'Azur, \\ BP 429
06304 Nice (France).}
\email{regimbau@obs-nice.fr}

\date{\today}
\begin{abstract}
We review the contribution of extra galactic
inspiralling double neutron stars, to the LISA astrophysical
gravitational wave foreground. 

Using recent fits of the star formation rate, 
we show that sources beyond
$z_*=0.005$ contribute to a truly continuous background, which may dominate the
LISA instrumental noise in the range $3 \simeq 10^{-4}$ -
$1 \times 10^{-2}$ Hz and overwhelm the galactic WD-WD confusion noise
at frequencies larger than $\nu_o \simeq 2\times 10^{-3}$. 

\end{abstract}


\maketitle

\section{\label{intro}Introduction}

Compact neutron star binaries are among the most promising sources
of gravitational waves.
At low frequencies, the continuous inspiral signal may be detectable
by the space antenna LISA, while ground based interferometers such as VIRGO
\cite{bra}, LIGO \cite{abr}, GEO \cite{hou} or TAMA \cite{kur},
are expected to detect the last few minutes prior coalescence, at
frequencies up to 1.4-1.6 kHz.

In a first paper \cite{dfp} (hereafter paper I), we have investigated the 
high frequency signal and its detection with the first generations of
ground based detectors.
Using new estimates of the mean merging rate in the local universe,
that account for the galactic star formation history derived directly
from observations and include the contribution of elliptical
galaxies, we predict a detection every 148 and 125 years in the volume 
probed by initial VIRGO and LIGO, and up to 6 detections
per year in their advanced configuration.

In a second paper \cite{reg06} (hereafter paper II), we used numerical simulations to estimate the
gravitational wave emission produced by the superposition of unresolved extra-galactic sources. 
As in paper I, we were interested in the few sixteen minutes before the last stable orbit is reached, 
when more than 96\% of the gravitational wave energy is released and when the frequency evolves in the range 10-1500 Hz, covered by
ground based interferometers.
We find that above the critical redshift $z_* = 0.23$, the sources
produce a truly continuous background, 
with a maximal gravitational density parameter of $\Omega_{gw} \simeq 1.1\times 10^{-9}$ around 670 Hz. 
The signal is below the sensitivity of the first generation of ground
based detectors but could be detectable by the next generation.
Correlating two coincident advanced-LIGO detectors or two third
generation EGO interferometers, we derived a S/N ratio of 0.5 and 10 respectively.

In this article, we extend our previous simulations to investigate the
stochastic background from the early low frequency inspiral phase.
Estimates of the emission from the various population of compact
binaries (see for instance \cite{eva,hils,ben,pos,nel,tim} for the galactic
contribution and \cite{sch,far,coo} for the extra-galactic
contribution) or from captures by supermassive black holes (\cite{bar}), which represent the main source of confusion noise for LISA, are of crucial interest
for the development of data analysis strategies.
The paper will be organized as follow. In II., we describe the
simulations of the DNS population, in III. the contribution of
inspiralling DNS to the stochastic background is calculated and
compared to the LISA instrumental noise and to the WD-WD galactic
foreground. Finally, in IV. the main conclusions are summarized.

\section{Simulations of the DNS population}

To generate a population of DNSs (each one characterized by its redshift
of formation and coalescence timescale) we follow the same procedure as
in paper II, to which the reader is referred for a more detailed description. 

The redshift of formation is randomly selected from the probability
distribution \cite{cow02}:
\begin{equation}
P_f(z_f)=\frac {R_{zf}(z_f)}{R_p}
\label{eq-proba_frate}
\end{equation}
constructed by normalizing the differential DNS formation rate,
\begin{equation}
R_{zf}(z_f)=\frac{dR_f(z_f)}{dz_f } 
\label{eq-frate1}
\end{equation}
The normalization factor in the denominator corresponds to the rate at
which massive binaries are formed in the considered redshift interval, e.g.,
\begin{equation}
R_{p}=\int_0^6(dR_f(z_f)/dz_f)dz_f,
\label{eq-Revents}
\end{equation}
which numerically gives $R_p = 0.044$ s$^{-1}$ in the two cosmic star formation rate adopted in
this paper.

The formation rate of massive binaries per redshift interval
eq.~\ref{eq-frate1} writes :
\begin{equation}
R_{zf}(z_f)=\lambda_{p} \frac{R^*_{f}(z_f)}{1+z_f}{\frac{{dV(z_f)}}{{dz_f}}}
\label{eq-frate}
\end{equation}
where $R^*_{f}(z)$ is the cosmic star formation rate (SFR)
expressed in M$_{\odot}$ Mpc$^{-3}$yr$^{-1}$ and $\lambda_{p}$ is the
mass fraction converted into DNS progenitors. 
Hereafter, rates per comoving volume will always be indicated
by the superscript '*', while rates with  indexes ``z$_f$" or ``z$_c$"
refer to differential rates per redshift interval, including all cosmological factors.
The (1+z) term in the denominator of eq.~\ref{eq-frate} corrects the star formation rate by time 
dilatation due to the cosmic expansion. 

The element of comoving volume is given by
\begin{equation}
dV = 4\pi r^2 \frac{c}{H_o} \frac{dz}{E(\Omega_i,z)}
\end{equation}
with
\begin{equation}
E(\Omega_i,z) =[\Omega_m(1+z)^3+\Omega_v]^{1/2}
\end{equation}
where $\Omega_m$ and $\Omega_v$ are respectively the present values of
the density parameters due to matter (baryonic and non-baryonic) and
vacuum, corresponding to a non-zero cosmological constant. 
Throughout this paper, the 737 flat cosmology \cite{rao}, with $\Omega_m$ = 0.30, $\Omega_v$ = 0.70 and Hubble parameter H$_0$ =
$70\,\mathrm{km\,s}^{-1}\mathrm{Mpc}^{-1}$ \cite{spe} is assumed.

Recent measurements of galaxy luminosity function in
the UV (SDSS, GALEX, COMBO17) and FIR wavelengths (Spitzer Space
Telescope), after dust corrections and normalization, 
allowed to refine the previous models of star formation history, up to
redshift $z\simeq 6$, with tight constraints at redshifts $z<1$.
In our computations, we consider the recent parametric fits  of the
form of \cite{cole}, provided
by \cite{hop}, constrained by the Super Kamiokande limit on the electron
antineutrino flux from past core-collapse supernovas. 
It is worth mentioning that the final results are not significantly
different if we adopt, as in paper II, the SFR given in \cite{por}.

Throughout this paper, we assume that the parameter
$\lambda_p$ does not change significantly with the redshift and can be
considered as a constant.
In fact, this term is the product of three other parameters, namely,
\begin{equation}
\lambda_{p}= \beta_{NS} f_b \lambda_{NS}
\end{equation}
where $\beta_{NS}$ is the fraction of binaries which remains
bounded after the second supernova event, $f_b$ is the fraction of massive binaries formed among all
stars and $\lambda_{NS}$ is the mass fraction of neutron star progenitors. 
From paper I we take $\beta_{NS}$ = 0.024 and $f_b$ = 0.136.

\cite{hop} investigated the effect of the initial mass
function (IMF) assumption on the normalization of the
SFR. They showed that top heavy IMFs are preferred to the traditional
Salpeter IMF \citep{salp} and their fits are optimized for IMFs of the form: 
\begin{equation}
\xi(m) \: \propto
\begin{array}{ll} 
\frac{m}{m_0}^{-1.5}    &   \hbox{  for } 0.1<m<m_0 \\
\frac{m}{m_0}^{-\gamma} &   \hbox{  for } m_0<m<100 \\
\end{array}
\label{eqn:imf}
\end{equation}
with a turnover below $m_0=1$ M$_{\odot}$, and normalized within the mass interval 0.1 - 100 M$_{\odot}$ such as $\int
m\xi(m)dm$ = 1.

In this paper, we assume $\gamma = 2.35$ (A modified Salpeter), but
taking $\gamma = 2.2$ (\cite{bal}) would not change the final results
significantly.
To be consistent with the adopted models of the SFR, 
we follow \cite{hop} and assume a minimal initial mass of
8 M$_{\odot}$ for NS progenitors, the upper mass being 40 M$_{\odot}$.
It results finally ,
$\lambda_{NS} =\int_8^{40}\xi(m)dm = 9 \times 10^{-3}$ M$_{\odot}^{-1}$.

The next step consists to estimate the redshift $z_b$ at which the progenitors
have already evolved and the system is now constituted by two neutron stars.
This moment fixes also the beginning of the inspiral phase. If $\tau_b$
($\approx 10^8$ yr) is the mean lifetime of the progenitors (average weighted by the 
IMF in the interval 8-40 M$_{\odot}$) then

\begin{equation}
z_b = z_f - H_0 \tau_b (1+z_f)E(z_f)
\label{eq-zb}
\end{equation}

Once the beginning of the inspiral phase is established, the redshift at which the coalescence
occurs should now be estimated. The duration of the inspiral depends on the orbital parameters
and the neutron star masses. The probability for a given DNS system to coalesce in a timescale
$\tau$ was initially derived by \cite{pa97} and confirmed by subsequent simulations \cite{vin,dfp} and is given by

\begin{equation}
P(\tau)=B / \tau
\label{eq-proba_tau}
\end{equation}
 
Simulations indicate a minimum coalescence timescale $\tau_0 = 2 \times 10^5$ yr and a considerable 
number of systems having coalescence timescales higher than the Hubble time. The normalized
probability in the range $2\times 10^5$ yr up to 20 Gyr implies $B=0.087$. Therefore,
the redshift $z_c$ at which the coalescence occurs is derived by solving the
equation
\begin{equation}
\tau = \tau(z_c,z_b)
\label{eq-tau_coal}
\end{equation} 
where  
\begin{equation}
\tau(a,b) = \int_{a}^{b} \frac{dz}{H_0(1+z)E(z)}
\label{eq-tau}
\end{equation} 
\section{The Gravitational Wave Background}

Compared with the previous study, slight complications arise when calculating the spectral properties of the signal. 
In paper II, we were interested in the signal emitted
between 10-1500 Hz, which duration in the source frame is short
enough, less than 1000 s, to be considered as a burst located at the redshift of coalescence. 
Now, as we are looking at inspiral phase in the LISA frequency window ($10^{-4}-0.1$ Hz), the evolution of the redshift
of emission with frequency must be taken into account. 
For the same reason, the critical redshift at
which the population of DNSs constitute a truly continuous background can't be
determined by simply solving the condition $D(z)>1$, where D is the duty
cycle defined as the ratio of the typical duration of a single
burst $\bar{\tau}$ to the average time interval between successive
events (see paper II).
Instead, we remove the brightest sources, lying in the close Universe,
where the distribution of galaxies is expected to be highly
anisotropic, 
in particular due to the strong concentration of galaxies in the
direction of the Virgo cluster and the Great Attractor (see paper I),
and where the density is well below the average value derived from the
SFR (\cite{pach06}).
In a second
step, we verify that the resulting background is continuous, by using
both Jaque Bera and Lilliefors gaussianity tests with significant
level $\alpha = 0.05$ over a set of 100 realizations. 

For each realization, the number of simulated DNSs corresponds to the
expected number of extragalactic DNSs observed today and is derived
as,
 \begin{equation}
N_p=\int_{0}^{6}N(z)dz
\label{eq-Np}
\end{equation}
The DNSs present at redshift z were formed at $z'_b \geq z$,
with a coalescence time larger than the cosmic time between $z'_b$ and
z.
Thus,
\begin{equation}
N(z)=\int_{z}^{6} R_{zc}(z') \eta(z') dz'
\end{equation}
where
\begin{equation}
\eta(z')= \int_{\tau_{min}}^{\tau_{max}} P(\tau) d\tau
\end{equation}
with $\tau_{max}$= 20 Gyr and $\tau_{min}$=Max($2\times 10^5$
yr;$\tau(z,z'_b)$), which gives $N_p \simeq 1.4 \times
10^6$.

We obtain that after $z_*=0.005$, corresponding to the distance beyond
the Virgo cluster ($\simeq 25$ Mpc), when the relative density
fluctuations are $< 0.5$ (\cite{pach06}), the background is continuous in the range
$10^{-4}-0.1$ Hz.
We consider also a more conservative threshold of $z_* = 0.02$,
corresponding to the distance beyond the Great Attractor ($\simeq 100$
Mpc), when the relative density fluctuations are $< 0.05$ and when the
Universe is expected to become homogeneous(\cite{pach06,wu96}).

The gravitational flux in the observer frame produced by a given DNS is:
\begin{equation}
f_{\nu _{o}}=\frac{1}{4\pi d_{L}^{2}}\frac{dE_{gw}}{d\nu }(1+z_e)
\label{eq-fluence}
\end{equation}

where $z_e$ is the redshift of emission, $d_{L}=(1+z_e)r$ the
distance luminosity, $r$ the proper distance, which depends on the adopted cosmology,
${dE_{gw}}/{d\nu}$ the gravitational spectral energy and
$\nu=(1+z_e)\nu _{o}$ the frequency of emission in the source frame.
 
In the quadripolar approximation and for a binary system with
masses $m_1$ and $m_2$ in a circular orbit: 
\begin{equation}
dE_{gw}/{d\nu} = K \nu^{-1/3}
\end{equation}
where the fact that the gravitational wave frequency is twice the
orbital frequency was taken into account. 
Then
\begin{equation}
K = \frac{(G \pi)^{2/3}}{3} \frac{m_1m_2}{(m_1+m_2)^{1/3}}
\end{equation}
Taking $m_1 = m_2 = 1.4$ M$_{\odot}$, one obtains $K=5.2 \times 10^{50}$ erg Hz$^{-2/3}$.

The redshift of emission is derived by solving the relation \cite{sch}:
\begin{equation}
\nu^{-8/3} = \nu_{LSO} + K' (\tau(z_b,z_e))
\end{equation} 
where 
\begin{equation}
K' = \frac{(256 \pi^{8/3} G^{5/3})}{5 c^5} 
\end{equation}
(or numerically $K'=67.83$ yr$^{-1}$),
and where the time spent between redshifts $z_e$ and $z_b$ is given by:
\begin{equation}
\tau (z_b,z_e)= \tau - \int_{z_c}^{z_e} \frac{dz}{(1+z)E(z)}
\end{equation} 

\subsection{Properties of the background}

The spectral properties of the stochastic background are characterized
by the dimensionless parameter \cite{fer99}:

\begin{equation}
\Omega _{gw}(\nu _{o})=\frac{1}{c^{3} \rho _{c}}{\nu _{o}}F_{\nu _{o}}
\end{equation}

where $\nu _{o}$ is the wave frequency in the observer frame and $\rho
_{c}$ the critical mass 
density needed to close the Universe, related to the Hubble parameter $H_{0}$ by,

\begin{equation}
\rho _{c}=\frac{3H_{o}^{2}}{8\pi G}
\end{equation}

$F_{\nu_o}$ is the gravitational wave flux at the 
observer frequency (given here in erg
cm$^{-2}$Hz$^{-1}$s$^{-1}$) 
$\nu_o$, integrated over all sources above the critical redshift $z_*$, namely
\begin{equation}
F_{\nu _{o}}=\int_{z_*}^{z_{\max }}f_{\nu _{o}}dR_c(z)
\label{eq-flux}
\end{equation} 

In our simulations, the integrated gravitational flux is calculated by summing 
individual fluences (eq.~\ref{eq-fluence}), scaled by the ratio between the total formation rate of progenitors
(eq.~\ref{eq-Revents}) and the number of simulated massive binaries, as:
\begin{equation}
F_{\nu_o}=\frac{R_{p}}{N_{p}}\sum_{i=1}^{N_{p}} f_{\nu_o}^i 
\end{equation}

\begin{figure}
\centering
\includegraphics[angle=270,width=0.9\columnwidth]{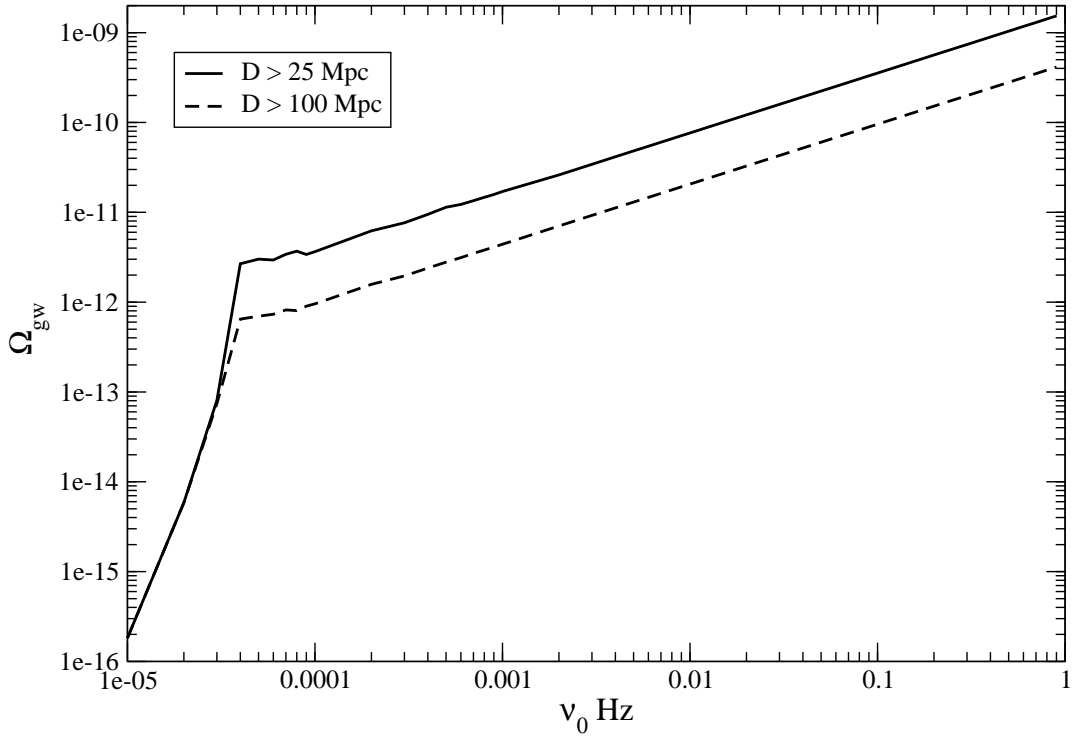}
\caption{
Spectrum of the expected gravitational energy density
  parameter $\Omega_{gw}$, corresponding to NS-NS inspirals
  occurring beyond $d\simeq 25$ Mpc (continuous curve) and
  $d \simeq 100$ Mpc (dashed curve). 
\label{fig-omega}} 
\end{figure}

Figure~\ref{fig-omega} shows the density parameter $\Omega_{gw}$ as a function of
the observed frequency averaged over 10 simulations.
The continuous line corresponds to sources located beyond the critical
redshift to have a continuous background, $z_*=0.005$ (or
$d\simeq25$ Mpc), while the
dashed line corresponds to the more conservative estimate, with a cutoff
at $z_*=0.02$ (or $d \simeq 100$ Mpc).
After a fast increase below $\simeq 5 \times 10^{-5}$ Hz, $\Omega_{gw}$ increases
as $\nu_{o}^{2/3}$ to reach $\Omega_{gw} \simeq 3 \times 10^{-10}$ at
0.1 Hz for $z_*=0.005$ ($\Omega_{gw}
\simeq 10^{-10}$ for $z_*=0.02$).
This is one order of magnitude larger than the previous predictions
by \citet{sch}.

\subsection{consequences for LISA}

\begin{figure}
\centering
\includegraphics[angle=270,width=0.9\columnwidth]{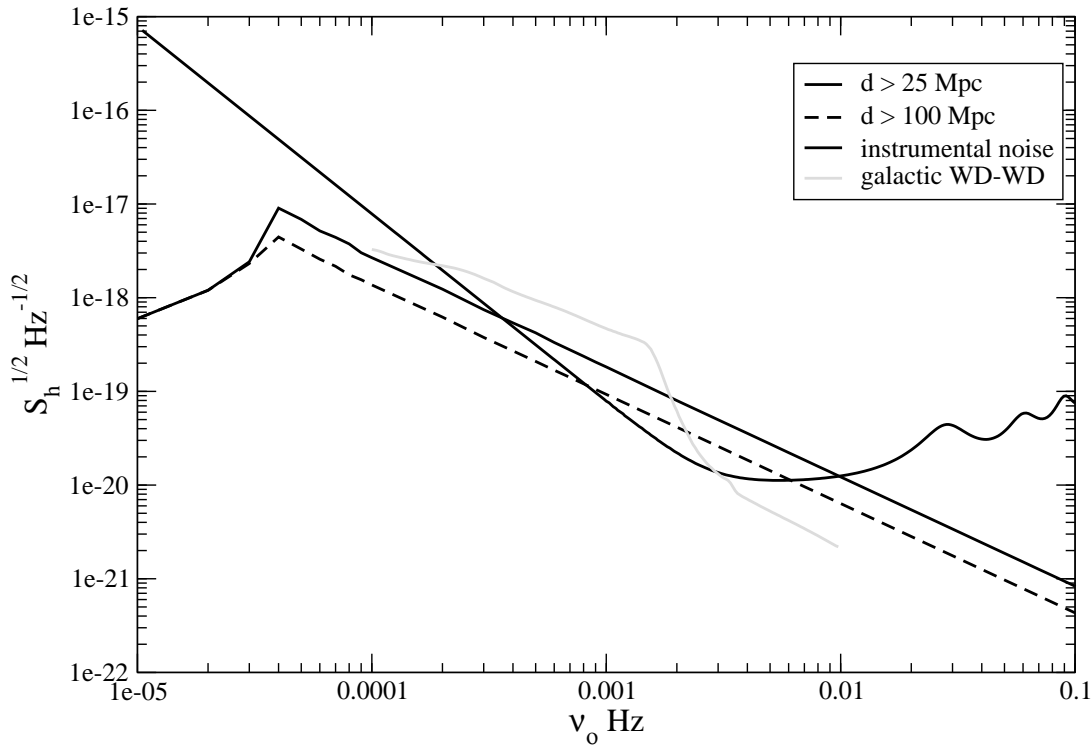}
\caption{
Gravitational strain in Hz$^{-1/2}$, corresponding to NS-NS inspirals
  occurring beyond $d\simeq 25$ Mpc (continuous curve) and $d \simeq 100$ Mpc
  (dashed curve), along with the LISA instrumental noise (black) and the
  galactic WD-WD foreground (light gray).   
\label{fig-Sh}} 
\end{figure}

Astrophysical backgrounds represent a
confusion noise for LISA, which spectral density is given by
\cite{bar}:
\begin{equation}
S_h^{back}(\nu_{o})=\frac{3H_{0}^2}{2 \pi^2} \frac{1}{\nu_{o}^3}\Omega_{\rm gw}(\nu_{o})
\end{equation}
Figure~\ref{fig-Sh} shows $S_h^{back}(\nu_{o})$ compared to the LISA
sensitivity  \cite{lar00} and to the contribution of unresolved
galactic WD-WD \cite{hils,lar00}.
The contribution from DNSs may dominate the LISA instrumental noise
between $\simeq 3\times 10^{-4}$ - $10^{-2}$ Hz for $z_*=0.005$
($\simeq 7\times 10^{-4}$ - $6\times 10^{-3}$ Hz for $z_*=0.02$) and the galactic WD-WD
confusion noise after $\simeq 2\times 10^{-3}$. 
However, the resulting reduction in the sensitivity should be less than a factor 4 and thus shouldn't affect
significantly signal detection. 

\begin{figure}
\centering
\includegraphics[angle=270,width=0.9\columnwidth]{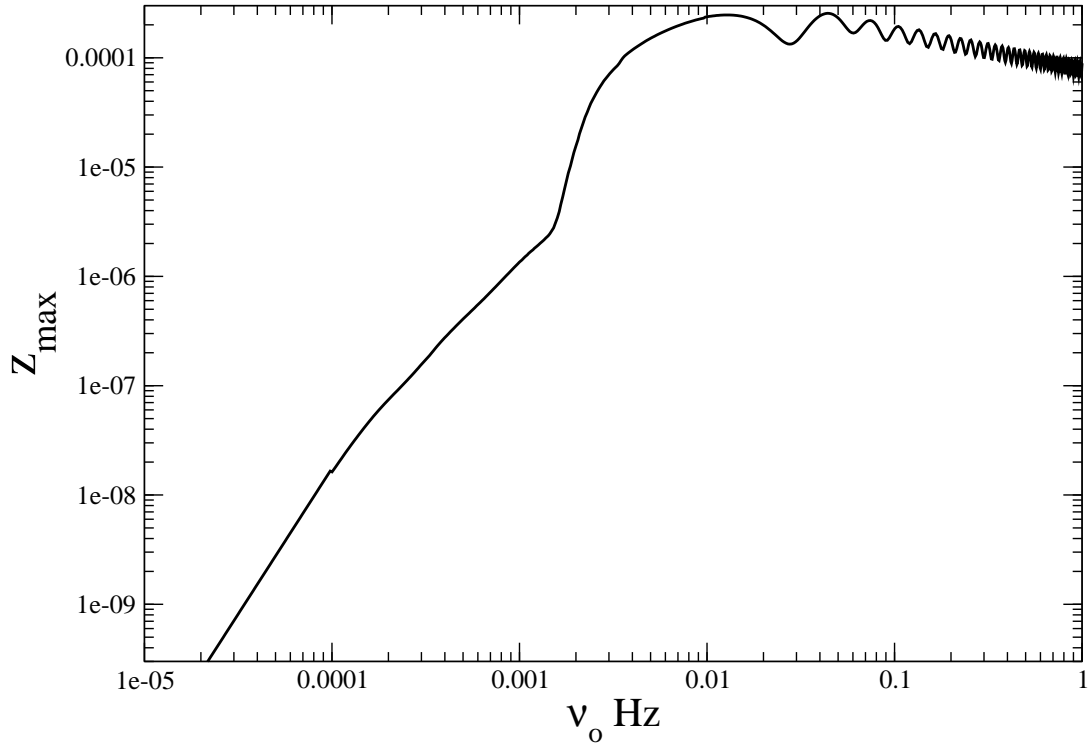}
\caption{
Maximal redshift  as a function of the observed frequency, at which a single DNS can be detected by LISA after
  one year of integration and with a signal-to-noise ratio $S/R=5$.
The galactic WD-WD foreground is taken into account in the LISA sensitivity.   
\label{fig-zmax}} 
\end{figure}

The critical redshift at which a single DNS can be resolved and
detected by LISA (Figure~\ref{fig-zmax}) is at least one order of
magnitude smaller ($z_{min}\simeq 3 \times 10^{-4}$ at frequencies
$\nu_o > 0.01$) than our threshold to
have a gaussian background $z_*=0.005$ (or $z_*=0.02$) .
Sources lying between the two regimes are responsible for a non
gaussian and non isotropic cosmic ``popcorn" noise (see \cite{cow06} for a recent
review) that will be investigated in a further work. 
This contribution, which statistical properties differ from those of
both the instrumental noise and the cosmological background, could be
detected and removed by new data analysis techniques currently under
investigation, such as the search for anisotropies \cite{all} that
can be used to create a map of the GW background \cite{cor,ball}, the
maximum likelihood statistic \cite{dra}, or methods based on the
Probability Event Horizon concept \cite{cow05}, which describes
the evolution, as a function of the observation time, of the cumulated signal throughout the Universe.
 
\section{Conclusions}
In this paper, we have modelled the contribution of extra galactic
inspiralling double neutron stars, to the to the astrophysical
gravitational wave foreground, in the range of sensitivity of the
space detector LISA. 

Using a recent fit of the star formation rate \cite{hop}, optimized
for a top heavy A modified Salpeter IMF, we show that sources beyond
$z_*=0.005$ (or $z_*=0.02$ for a more conservative estimate)
constitute a truly continuous background.
We find that the density parameter $\Omega_{gw}$, after a fast increase below $\simeq 5 \times 10^{-5}$ Hz,
increases as $\nu_{o}^{2/3}$ to reach $\Omega_{gw} \simeq 3\times 10^{-10}$ ($\Omega_{gw}
\simeq 10^{-10}$) at 0.1 Hz, which is one order of magnitude above the
previous estimate of \cite{sch}.
As a result, the signal may dominate the
LISA instrumental noise in the range $\simeq 3\times 10^{-4}$ -
$10^{-2}$ Hz ($\simeq 7\times 10^{-4}$ - $6\times 10^{-3}$) and overwhelm the galactic WD-WD confusion noise
at frequencies larger than $\nu_o \simeq 2\times 10^{-3}$. 
Sources located closer than $z_*$, but still too far to be detected by
LISA, constitute a non gaussian and non isotropic cosmic ``popcorn"
noise, which will be investigated in a further work. 

{\bf Acknowledgement}

The author thanks J.A. de Freitas Pacheco and A. Spallicci for usefull
discussions, which have improved the first versions of the paper.

\end{document}